\documentclass[epsf]{aa}

\usepackage{times}
\usepackage{graphics}

\begin{document}

\title{
{\bfseries{\itshape ORFEUS\,II} echelle spectra :\\
On the H$_2$/CO ratio in LMC gas towards LH\,10
}}
\author{P. Richter\inst{1}
\and K.S. de Boer\inst{1}
\and D.J. Bomans\inst{2}
\and Y.-N. Chin\inst{3,}\inst{4}
\and A. Heithausen\inst{4}
\and J. Koornneef\inst{5}
}

\institute{
Sternwarte, Universit\"at Bonn, Auf dem H\"ugel 71, D-53121 Bonn, Germany
\and
Astronomisches Institut, Ruhr-Universit\"at Bochum, Postfach 102148,       
        D-44780 Bochum, Germany
\and
Institute of Astronomy and Astrophysics, Academia Sinica,
   P.O.Box 1-87, Nankang, 11529 Taipei, Taiwan
\and
Radioastronomisches Institut, Universit\"at Bonn, Auf dem H\"ugel 71, 
        D-53121 Bonn, Germany
\and
Kapteyn Institute, Postbus 800, NL-9700AV Groningen, the Netherlands
}

\date{Received June 10 1999 / Accepted xxx 1999}

\thesaurus{03.19.2, 08.09.2\,LH\,10:3120, 09.13.2, 11.09.4, 11.13.1, 13.21.3} 

\offprints{prichter@astro.uni-bonn.de}

\maketitle

\markboth{P. Richter et al., On the H$_2$/CO ratio in LMC gas towards LH\,10}
        {P. Richter et al., On the H$_2$/CO ratio in LMC gas towards LH\,10}

\begin{abstract}
{\it ORFEUS} far UV echelle spectra have been used
to investigate H\,{\sc i}, H$_2$ and CO absorption lines 
along the line of sight towards LH 10:3120 in the LMC,
extending the study presented by de\,Boer et al. (1998).
While H$_2$ absorption is clearly visible, no CO
absorption at LMC velocities is detected, but an upper limit
of $N($CO$) \le 3.3 \times 10^{13} $ cm$^{-2}$ 
for the CO column density is derived for the
C-X band near 1088 \AA.
The detected H$_2$ absorption features 
were used to determine a limit
for the H$_2$/CO ratio of $N$(H$_2$)/$N$($^{12}$CO)
$ \ge  2.0 \times 10^5$ for the LMC gas along 
this individual line of sight. 
Generally, the fraction of gas in molecular 
form in the LMC is low compared to interstellar gas in the
Milky Way with the same total gas quantity. 
We compare the absorption spectroscopy
with {\it SEST} CO emission line measurements.
It is found that most of the CO emission comes from gas
behind LH 10:3120.
We discuss our results in view of the possible
scenario, in which the low dust content 
limits the amount of molecular hydrogen in 
the diffuse LMC gas. 

\keywords{Space vehicles - ISM: molecules - Galaxies: ISM - 
        Magellanic Clouds: LMC - Stars: individual: LH 10:3120
        - Ultraviolet: ISM}

\end{abstract}

\section{Introduction}
Molecular hydrogen (H$_2$) is by far the most abundant molecule in the interstellar
medium, followed by carbon monoxide (CO) with a significantly smaller amount.
The symmetry of H$_2$ results in
the absence of a permanent dipole momentum, which makes the 
dominant interstellar molecule very difficult to detect.
Warm H$_2$ gas can be observed in near IR emission, but 
the largest amount of H$_2$ is located in cold gas, 
which is measurable in FUV absorption along 
individual lines of sight against bright UV background sources. 
Thus the {\it overall} amount of H$_2$ in the interstellar medium
can not be measured directly.  
The amount of H$_2$ often is
estimated from the abundance of CO, 
which can easily be determined from radio emission
lines. For this method it is necessary to know the
conversion factor $X$ between H$_2$ column density and 
the velocity-integrated $^{12}$CO emission temperature,
defined as 
$X(^{12}$CO$)=N($H$_2) / W(^{12}$CO$)$ (e.g. Heithausen \& Mebold 1989).

The $X$ factor has been determined for the Milky Way with some degree
of accuracy, using various (indirect) methods (Bloemen et al. 1986).
Obviously, it is important to determine the conversion factors in the
most nearby lower-metallicity systems (the Magellanic Clouds), too.
Cohen et al. (1988) obtained a value which is 6 times as large as in our
Galaxy and Rubio et al. (1993) noticed the possible 
dependence of $X$ on the cloud size.
Both results could be affected by the limited angular resolution.
Recently, Chin et al. (1997, 1998) estimated the conversion factors
in selected regions in the LMC and SMC. A review of molecules and the 
$X$ factor in the Magellanic Clouds is available from Israel (1997).

An important new aspect for the knowledge of interstellar molecular
gas in galaxies with chemical abundances different from those
of the Milky Way
is the 
determination of {\it real} abundance ratios of the H$_2$ and CO molecules
via UV absorption spectroscopy. 
This is also the most elegant and safe way to decrease the uncertainties in the
measurements of the $X$ factor in those systems.
The {\it Copernicus} satellite could measure the
relation between H$_2$ and CO for the diffuse interstellar gas in the Milky Way
(Federman et al. 1980). 
However, every individual H$_2$/CO ratio definitely is influenced
by the metallicity, by the local radiation field and
by the amount of dust in the observed complex.
So far, H$_2$/CO ratios 
from absorption spectroscopy 
could not be measured outside the Milky Way.
The sensitvity of {\it Copernicus} was not
high enough while
the {\it IUE} satellite as 
well as the {\it HST}
do not cover the wavelength range below 1200 \AA,
where H$_2$ is found in absorption.
{\it ORFEUS} is the first instrument able to 
investigate the cold molecular gas in the Magellanic Clouds 
via absorption spectroscopy.

Using {\it ORFEUS} data we measured H$_2$ absorption profiles
toward one LMC (de Boer et al. 1998, hereafter Paper\,I) and one SMC (Richter
et al. 1998) star. 
Among the observed Magellanic Cloud lines of sight only the
one to the star 3120 (Parker et al. 1992) in the association 
LH 10 in the LMC shows a column densitiy in H$_2$
high enough to expect the presence of CO.
The {\it ORFEUS}
spectrum of LH 10:3120 therefore
provides the very first opportunity to determine a
H$_2$/CO ratio for LMC gas directly.
 
\section{{\bfseries{\itshape ORFEUS}} absorption line measurements}

\subsection{Instrument description}

The spectroscopic data were obtained with the 
Heidelberg-T\"ubingen echelle spectrometer during
the {\it ORFEUS II} mission on the {\it ASTRO-SPAS}
space shuttle mission between November and
December 1996.
{\it ORFEUS} (Orbiting and Retrievable Far
and Extreme Ultraviolet Spectrometer) consists of
a 1\,m normal incidence mirror with a focal length of 2.4\,m.
The echelle operated in the diffraction orders 40 to 
61 in the wavelength range from 900 to 1400 \AA.
The echelle spectrometer was designed to achieve
a spectral resolution of $\lambda / \Delta \lambda =
10^4$ in an entrance aperture of $10 ''$ 
(Appenzeller et al. 1988), but turned out
to be more accurate ($\simeq 1.2 \times 10^4$)
during the measurements (Barnstedt et al. 1999).
At 1000 \AA\, this value corresponds to a resolution
of 83 m\AA\, or 25 km\,s$^{-1}$.

The echelle detector is a photon counting 
microchannel plate detector with 1024 by 512 pixels 
in an active area of $44 \times 44$\,mm.
A detailed description of the instrument is given
by Barnstedt et al. (1999).

\subsection{Observations} 

The target star for our observation was star 3120,
located in the association LH\,10 in the north-western
part of the LMC in the N\,11 superbubble complex (see Paper\,I).
Basic properties of the background star (Parker et al. 1992) have been
summarized in Table\,1. The star has been observed
in 3 pointings with a total observing time of $\sim 6000$\,s.

\begin{table}[h]
\caption[]{Basic properties of the target star in the LMC}
\begin{tabular}{lccccc}
\hline
Object & $V$     & $E(B-V)$ & Spectral & $l$ & $b$\\
       & [mag] &  [mag]   &  type     \\
\hline
\rule [-2mm]{0mm}{6mm}{LH\,10:3120} & 12.80 & 0.17 & O5.5\,V & $277.2$ & $-36.1$\\ 
\hline
\end{tabular}
\noindent
\end{table}

\subsection{Data reduction}

The basic data reduction, such as the correction
of the blaze function, the wavelength calibration and
the background substraction, has been performed by the 
{\it ORFEUS} team in T\"ubingen. 
The mean accuracy for the wavelength calibration is
$\le 50$ m\AA\, and thus better than the
spectral resolution of the instrument (Barnstedt
et al. 1999). Moreover, the wavelength scale 
has been corrected for the orbital movement of
the satellite and the Earth's movement.
The background substraction, as described
in Barnstedt et al. (1999), has taken into account
contributions from the intrinsic background of the
detector, from straylight and from particle 
events related to the South Atlantic Anomaly.
For the flux calibration they give an internal 
accuracy of $\pm 10 \%$.

Because of the low countrate in the spectrum of LH\,10:3120 
the data has been
filtered by a de-nosing algorithm based on the powerful wavelet
transformation, as described by Fligge \& Solanki (1997).
In principle, the noise in the data is consistent with that 
expected from photon statistics.
The de-noising procedure increases the signal-to-noise 
ratio (S/N) by a factor of $\sim \sqrt{3}$, 
at the cost
of a slight decrease in the spectral resolution. 
However, since the target was
well centered within the diaphragm, the intrinsic
spectral resolution is better than $10^4$
(see section 2.1).
With respect to the pixel size the 
resulting spectral resolution after filtering turned
out to be $\sim 30$ km\,s$^{-1}$.

\begin{table}[t]
\caption[]{H$_2$ equivalent widths for the Lyman band in the LMC gas towards LH\,10:3120}
\begin{tabular}{lcccc}
\hline
\hline
\rule [-2mm]{0mm}{6mm}{Line} & $\lambda$ [\AA] & $f$ & $W_{\lambda}$ [m\AA] & Error [m\AA]\\
\hline
\hline
\multicolumn{5}{c}{\rule [-2mm]{0mm}{6mm}{$J=0$, $g_J=1$, $E_J=0$ eV, log $N=18.7 \pm 0.2$ cm$^{-2}$}}\\
\hline
\vspace{0.02cm}\\
R(0),0-0 &  $1108.128$ & $0.00173$ & $186$ & $31$ \\
R(0),1-0 &  $1092.194$ & $0.00596$ & $244$ & $40$ \\
\vspace{0.12cm}\\
\hline
\multicolumn{5}{c}{\rule [-2mm]{0mm}{6mm}{$J=1$, $g_J=9$, $E_J=0.01469$ eV, log $N=18.1 \pm 0.1$ cm$^{-2}$}}\\
\hline
\vspace{0.02cm}\\
R(1),0-0 &  $1108.634$ & $0.00117$ & $104$ & $23$ \\
R(1),1-0 &  $1092.732$ & $0.00403$ & $154$ & $25$ \\
R(1),2-0 &  $1077.698$ & $0.00809$ & $191$ & $39$ \\
P(1),2-0 &  $1078.925$ & $0.00385$ & $137$ & $32$ \\
P(1),4-0 &  $1051.031$ & $0.00403$ & $154$ & $25$ \\
\vspace{0.02cm}\\
\hline
\multicolumn{5}{c}{\rule [-2mm]{0mm}{6mm}{$J=2$, $g_J=5$, $E_J=0.04394$ eV, log $N=17.7 \pm 0.2$ cm$^{-2}$}}\\
\hline
\vspace{0.02cm}\\
R(2),0-0 & $1110.120$ & $0.00107$ & $108$ & $25$ \\
R(2),2-0 & $1079.226$ & $0.00739$ & $124$ & $28$ \\
R(2),5-0 & $1038.690$ & $0.01700$ & $155$ & $33$ \\
\vspace{0.02cm}\\
\hline
\multicolumn{5}{c}{\rule [-2mm]{0mm}{6mm}{$J=3$, $g_J=21$, $E_J=0.08747$ eV, log $N=17.5 \pm 0.4$ cm$^{-2}$}}\\
\hline
\vspace{0.02cm}\\
R(3),4-0 & $1053.976$ & $0.01420$ & $123$ & $26$ \\
R(3),5-0 & $1041.156$ & $0.01640$ & $139$ & $32$ \\
P(3),5-0 & $1043.498$ & $0.01040$ & $129$ & $28$ \\
\vspace{0.02cm}\\
\hline
\multicolumn{5}{c}{\rule [-2mm]{0mm}{6mm}{$J=4$, $g_J=9$, $E_J=0.14491$ eV, log $N=16.7 \pm 0.4$ cm$^{-2}$}}\\
\hline
\vspace{0.02cm}\\
R(4),2-0 & $1085.144$ & $0.00700$ & $94$ & $31$ \\
P(4),2-0 & $1088.794$ & $0.00487$ & $83$ & $25$ \\
P(4),4-0 & $1060.580$ & $0.00930$ & $87$ & $28$ \\
\vspace{0.02cm}\\
\hline
\multicolumn{5}{c}{\rule [-2mm]{0mm}{6mm}{Total H$_2$: log $N=18.8 \pm 0.2$ cm$^{-2}$}}\\
\hline
\end{tabular}
\noindent
\end{table}

\subsection{H$_2$ equivalent width measurements}

%fig1
\begin{figure*}[t]
\resizebox{13cm}{!}{\includegraphics{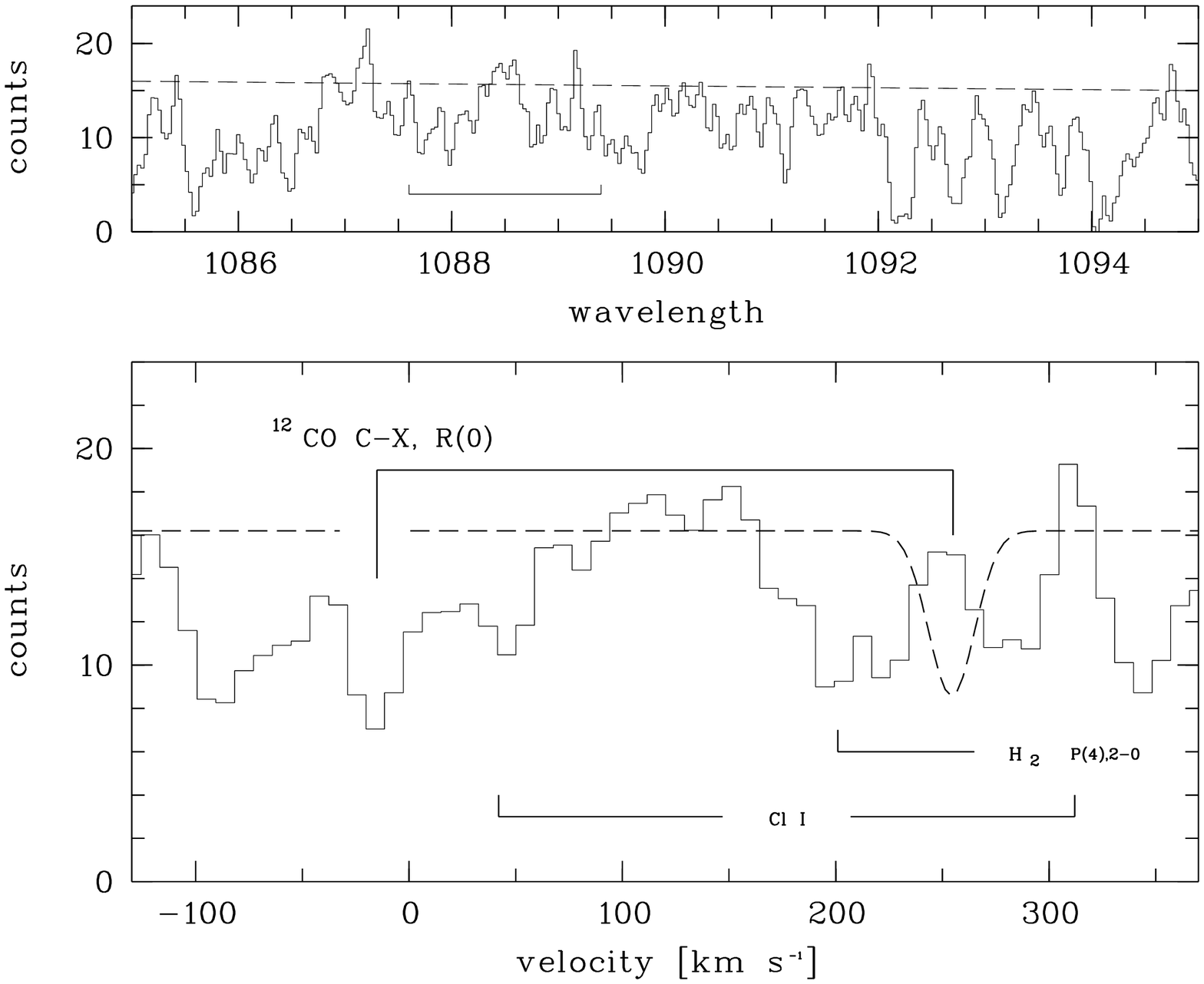}}
\hfill
\parbox[b]{40mm}{
\caption[]{
The spectrum near the $^{12}$CO  R(0),0-0 line in the
C-X band at 1087.87 \AA.\\
{\it Top panel:} The wavelength range between 1085 and 1095 \AA\, plotted in
$\lambda$-units. The dashed line indicates the adopted continuum level. 
The marked area is shown enlarged in the the lower panel.\\
{\it Bottom panel:}
The R(0),0-0 line plotted in the velocity scale (LSR).
The solid lines indicate the velocties at $-15$ km\,s$^{-1}$ (Milky Way)
and $+255$ km\,s$^{-1}$ (LMC). Galactic CO absorption
is visible, but no absorption structure
is present at the LMC velocity, as indicated by the
dashed line showing how a 40 m\AA\, strong CO line
at $+255$ km\,s$^{-1}$ would look like. 
Other possible absorption structures are identified 
}}
\end{figure*}

More than 100 H$_2$ transitions from the Lyman
and Werner bands are located
in the wavelength range between 915 and 1140 \AA.
For the further analysis we took wavelengths and oscillator
strengths from the list of Morton \& Dinerstein (1976).
In general, the strongest H$_2$ lines are located in the
range below 1000\,\AA.
Unfortunately, this is the part of the spectrum
where the decreasing sensitivity of the instrument as
well as the increasing UV extinction leads to 
extremely low countrates, so that we had to exclude 
this wavlength range from the further analysis.
Another problem is that the large number of transitions
in combination with the expected complex line-of-sight
structure let most of the H$_2$ lines overlap with
other H$_2$ absorption components. In addition, 
many atomic transitions, as compiled by 
Morton (1991), contribute to the
confused absorption structure in some parts of
the spectrum.
 
The low S/N ratio does not allow multi-component fits
for the absorption features so that decompositions
were not possible.
Despite these restrictions we found 16 H$_2$
absorption lines at $v_{\rm LSR} = +255$ km\,s$^{-1}$ 
from the lowest 5 rotational states, 
essentially free of lineblends in the velocity range between 
$+220$ and $+300$ km\,s$^{-1}$,
for which 
an analysis of line strengths was possible
\footnote{We note
at this point that the velocities presented
in Paper\,I were
labeled (by mistake) as LSR velocties while heliocentric velocities
were meant. We here correct to LSR velocities
with $v_{\rm LSR} = v_{\rm helioc.} - 15$ km\,s$^{-1}$}. 
For these lines, the equivalent widths were measured by
trapezium and (in some cases) by gaussian fits. With respect to the low 
countrate any more
sophisticated method would not lead to
a higher accuracy. We measured the equivalent width of each line 3 times
independently and used the mean value for the 
further evaluation.

The formal uncertainties in the equivalent
widths are basically due to photon statistics and
due to the error for the
choice of the continuum.
The low count rate and the resulting 
low S/N ratio causes 
uncertainties in the determination of the
continuum level as well as in the 
analysis of the line strengths themself.
However, the choice of the continuum 
is in particular critical for weaker absorption
structures, but our selection contains basically 
strong absorption lines. 
We determined the uncertainty for each 
line using the following procedure :
we fitted (by eye) the continuum 
around the absorption line and (when
necessary) normalized the continuum
level. We fitted a maximum
and a minimum continuum in order
to give an upper and a lower limit
for its uncertainty, including the error range
for the background substraction.
In addition, we obtained the formal uncertainties in 
the equivalent widths (for a fixed continuum) 
using the algorithm of Jenkins et al. (1973).
The total uncertainty for the equivalent width 
measurements then   
was calculated by taking both contributions
into account. On the average, the two error
sources contribute approximately equal to the total 
uncertainty.

All H$_2$ equivalent widths 
and their errors have been compiled in
Table 2. Examples for absorption line profiles
can be found in Paper\,I.

\subsection{H$_{2}$ column densities}

For the further investigation the 
equivalent widths were used for a standard 
curve-of-growth analysis for the LMC gas.
We find that the best fit for the curve of
growth has a $b$-value of 
$5$ km\,s$^{-1}$ (see Paper I). For the lower rotational
states the $b$-value is not critical, since
their lines are strong and located near the damping part
of the curve of growth.
Only for $J=4$ a 
different $b$-value would lead to a significant change in
the column density for this state.
The individual column densities $N(J)$ derived with
this method are presented in Table 2.
They range from
$N$(H$_{2})=5.0 \times 10^{16} $ cm$^{-2}$ for $J=4$
to $N$(H$_{2})=4.5 \times 10^{18} $ cm$^{-2}$
for $J=0$. 
Fixing the $b$-value at $5$ km\,s$^{-1}$ the
uncertainty for the individual column densities $N(J)$
has been derived by shifting the data points
on the curve of growth within their
individual error bars to the largest and
smallest possible column density. From
these values we then determined a mean deviation
to the best fit of $N(J)$, as given
in Table 2.
Summing over all $N(J)$, we find a total H$_2$ column density of
$N$(H$_{2})_{\rm total} = 6.6 ^{+3.4} _{-2.6} \times 10^{18} $ cm$^{-2}$.
The empirical curve of growth for all states
is shown in Fig.\,2 of Paper\,I.

\subsection{H$_{2}$ rotational excitation}

The determination and interpretation of the rotational
excitation of the H$_2$ gas has been described in 
Paper\,I. For $J=0,1$ we find an 
excitation temperature of $\le 50$ K from
a fit of a Boltzmann distribution, representing
the kinetic gas temperature. For $J=2,3$ and $4$  the
equivalent Boltzmann temperature is $470$ K, 
most likely indicating moderate UV pumping.
 
\subsection{CO absorption}

We investigated the amount of carbon monoxide (CO)
in the {\it ORFEUS} spectrum of LH 10:3120 by
looking for absorption in the strongest of
the $^{12}$CO absorption bands, the
C-X band near 1088 \AA.
Other CO bands in the wavelength range of {\it ORFEUS},
such as the
E-X band near 1076 \AA, show blendings at LMC
velocities or
are too weak in comparison to the C-X band.
We used wavelengths
and oscillator strengths for the CO transitions
from the list of Morton \& Noreau (1994).
Fig\,1 shows the spectral region of the C-X R(0), 0-0
transition.
Due to the low S/N ratio
and due to a probable blending from Cl\,{\sc i} 
at 1088.06 \AA\ the absorption structure looks
rather complex.
Moreover, the H$_2$ P(4), 2-0 line with its galactic component
at 1088.79 \AA\ overlaps the CO absorption
in the plotted region, as indicated in Fig.\,1.
Galactic CO absorption near
$-15$ km\,s$^{-1}$ is present
\footnote{Galactic CO absorption is also visible in
the weaker E-X band near 1076 \AA, so that we conclude
that the absorption structure at $-15$ km\,s$^{-1}$ in
Fig.\,1 doubtless is due to galactic CO absorption, too. 
The analysis
of the molecular absorption lines related to Milky Way
gas will be presented elsewhere}
but there is definitely no absorption 
visible at $+255$ km\,s$^{-1}$, where the
molecular hydrogen in the LMC gas has its absorption (see Fig.\,1)
and where possible CO absorption from LMC gas should be visible.
Obviously, the amount
of CO in the LMC gas is to small to be
detectable in the {\it ORFEUS} spectrum of LH\,10:3120.
Because of the low count rate noise peaks have
equivalent widths comparable with those of real
absorption lines.  
Typical noise features are visible near $+280$ and $+340$ km\,s$^{-1}$.

In the following, we derive an upper limit for the $^{12}$CO
column density by estimating the detection
limit for the C-X band.
The dominant amount of the CO populates the
two ground states with $J=0,1$ (Morton \& Noreau 1994).
For the C-X band, the transitions for these states, $R(0)$, $R(1)$ and
$P(1)$, can not be resolved
with the resolution of {\it ORFEUS}. 
For a rotational temperature of $\sim$5 K for
the CO molecules (as typically found in molecular gas in
the Milky Way), the two lowest rotational states
($J=0,1$) are approximately equally populated
(Morton \& Noreau 1994).
Under this assumption we can
sum over the individual oscillator
strengths $f$ of  $R(0)$, $R(1)$ and $P(1)$ and find
a total value of $f_{\rm 0,1} = 0.235$.
Based on the noise present and the resulting
continuum uncertainty, we place a detection limit 
of $\simeq 40$ m\AA\,
in the {\it ORFEUS} spectrum
near 1088 \AA\, using the 
procedure described in 
Sect.\,2.4. This value fits well
with the actual upper limit of the strength of a noise peak, 
as measured in the 
selected wavelength range.
In Fig.\,1 a $40$ m\AA\, strong absorption structure 
at $+255$ km\,s$^{-1}$ is shown as illustration.
We find an upper limit for the CO column density in the LMC gas
of $N($CO$) \le 3.3 \times 10^{13} $ cm$^{-2}$ based on
the $40$ m\AA\, detection limit and the curve
of growth with $b=5$ km\,s$^{-1}$.
This value is valid under the assumption that the H$_2$
and the CO are located in the same clouds.
In that case, the upper limit in $N($CO$)$ for each cloud scales
downward linearly with $W_{\lambda}$ provided that the
used $b$-value is scaled down by the same factor.
For multiple identical clouds the limit for
$N($CO$)_{\rm total}$ would be the same as the one
derived above.

%fig2
\begin{figure}
\resizebox{1.0 \hsize}{!}{\includegraphics{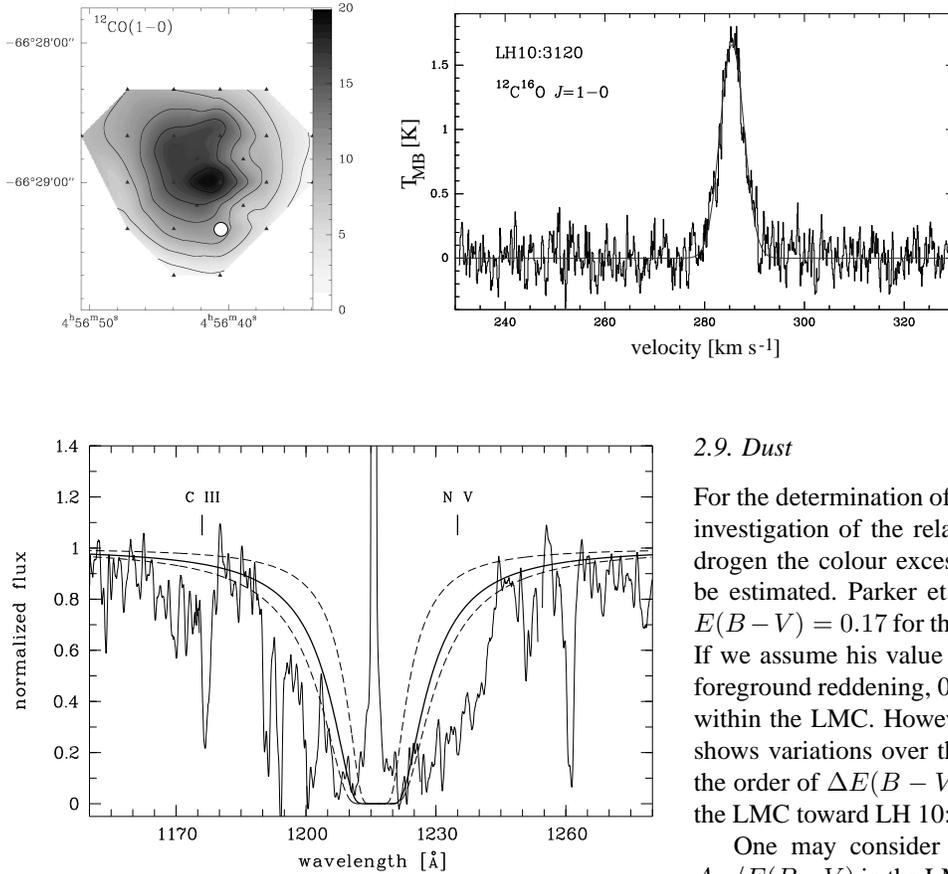}}
\caption[]{
The H\,{\sc i} Ly\,$\alpha$ absorption, fitted
by a Voigt profile (solid line) with contributions of 
$N$(H\,{\sc i})$ = 5.0 \times 10^{20}$ cm$^{-2}$
from galactic gas at
$-15$ km\,s$^{-1}$ and 
$N$(H\,{\sc i})$ = 2.0 \times 10^{21}$ cm$^{-2}$ from LMC gas at $+255$ km\,s$^{-1}$.
For comparison, the dashed lines show fits with H\,{\sc i} column densities
of $0.8$ and $3.0 \times 10^{21}$ cm$^{-2}$ for the LMC component.
Stellar absorption features in the damping wings
of Ly\,$\alpha$ are identified above the spectrum.
Geocoronal H\,{\sc i} produces the strong emission
in the centre of the absorption structure
}
\end{figure}

%fig3
\begin{figure*}
\resizebox{13cm}{!}{\includegraphics{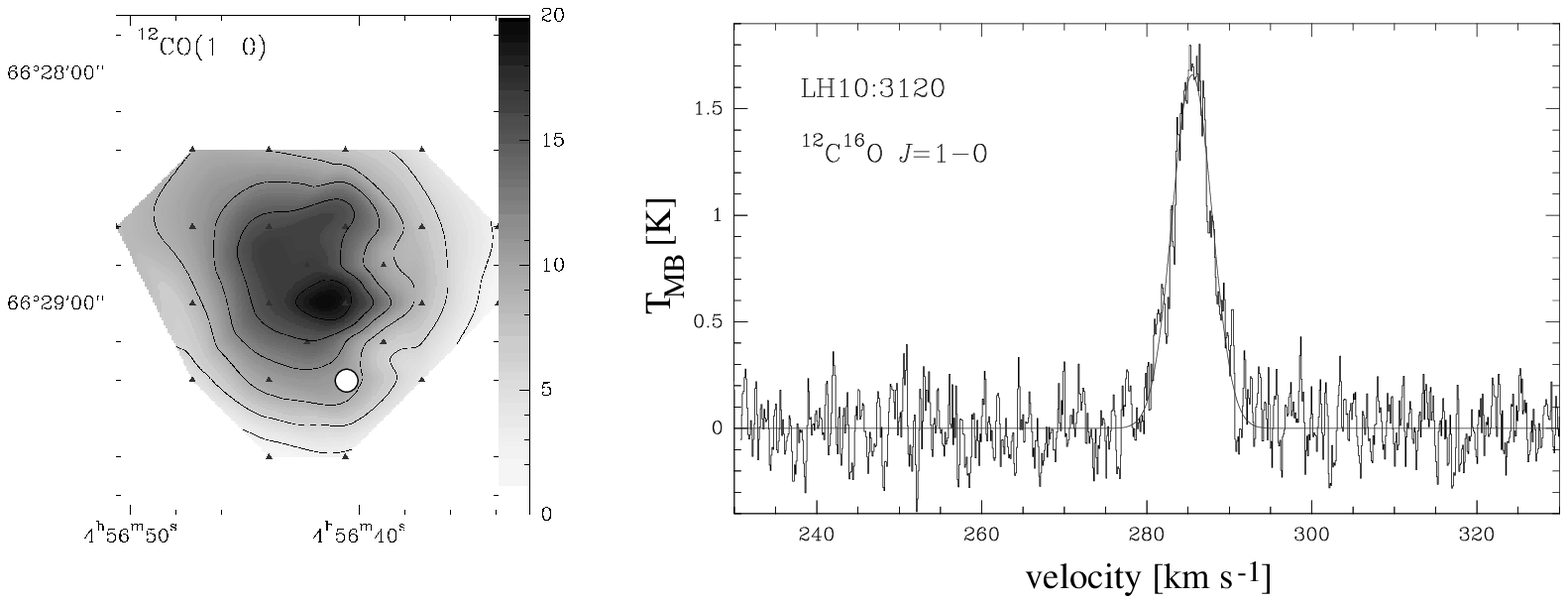}}
\hfill
\parbox[b]{40mm}{
\caption[]{
Radio measurements of $^{12}$CO are presented.\\
{\it Left panel:} Contour map of $^{12}$CO in the direction towards
LH 10:3120 (marked with the open circle). The amount of CO
increases to the North, where the geometrical centre of LH 10 is
located.\\
{\it Right panel:} The spectrum of $^{12}$CO shows the
maximum intensity near $+286$ km\,s$^{-1}$ (LSR)
}}
\end{figure*}

\subsection{Neutral Hydrogen}

There are no accurate 21\,cm data available for the 
sight-line to LH 10:3120, but  
the H\,{\sc i} column density in this direction
can be determined by
fitting the damping wings of the strong Ly\,$\alpha$ absorption line
in the {\it ORFEUS} spectrum.
Obviously, the Ly\,$\alpha$ line (Fig.\,2)
is blended by the stellar N\,{\sc v} line near 1240 \AA\ and
by some strong interstellar absorption lines near 1190 \AA\ (e.g. Si\,{\sc ii}).
In addition, the noise also disturbs the shape of
the profile.
No P\,Cygni profile is visible for the stellar N\,{\sc v}, as seen in
spectra of other early type stars.
Since the right wing of the Ly\,$\alpha$ absorption is completely
blended by the stellar N\,{\sc v} line, we fit the left wing of 
Ly\,$\alpha$ with a theoretical Voigt profile, using
oscillator strengths and damping parameters from
the list of Morton (1991).  

The H\,{\sc i} absorption from galactic foreground gas 
in direction to the LMC is known to be 
$\sim 3-4 \times 10^{20}$ cm$^{-2}$ (McGee et al. 1983).
Moreover, from the galactic S\,{\sc ii} abundance
in our spectrum
($N$(S\,{\sc ii})$ \le 8.0 \times 10^{15}$ cm$^{-2}$) 
in combination with the known ratio
[S\,{\sc ii}/H]$_{\rm MW} = -4.8$ dex (de Boer et al. 1987) 
we place un upper limit of $5.0 \times 10^{20}$ cm$^{-2}$
for galactic foreground absorption in H\,{\sc i} in front of LH\,10:3120.
For the profile fit we fixed the galactic component ($-15$ km\,s$^{-1}$)
at the latter column density and varied the LMC component at $+255$ km\,s$^{-1}$
in order to reproduce the observed absorption structure.
We find the best agreement with a H\,{\sc i} column density
of $N$(H\,{\sc i})$ = 2.0 \times 10^{21}$ cm$^{-2}$ for the
LMC gas, as shown in Fig.\,2 (solid line).
The uncertainty in this value is basically determined by the range
of possible profile fits and not so much due to the uncertainty in the
galactic foreground absorption.
Varying the H\,{\sc i} column density for the LMC component
we find a reasonable agreement for profile fits in a range between    
$0.8$ and 3.0 $\times 10^{21}$ cm$^{-2}$, in Fig.\,2 shown
as dashed lines.

\subsection{Dust}

For the determination of the gas-to-dust ratio as well
as for the investigation of the relation between
dust and molecular hydrogen the colour excess
$E(B-V)$ {\it within} the LMC has to be estimated.
Parker et al. (1992) give a minimum value of
$E(B-V)=0.17$ for the total reddening towards
LH 10:3120. If we assume his value of $E(B-V)=0.05$ as the
galactic foreground reddening, 0.12 mag remains
for the colour excess within the LMC.
However, the galactic foreground reddening shows
variations over the field of the LMC (Bessel 1990) of the
order of $\Delta E(B-V)=0.05$, so that the colour excess in
the LMC toward LH 10:3120 is uncertain to the same degree.

One may consider that the selective extinction $R_V = A_V/E(B-V)$
in the LMC is different from that in the Galaxy. Research in the
past has shown that the difference between $R_{V,{\rm LMC}}$ and
$R_{V,{\rm MW}}$ is negligible with regard to the expected error
for $E(B-V)_{\rm LMC}$ (Koornneef 1982; Cardelli et al. 1989).
For the further discussion we will stick to the Galactic 
$R_V$-value. The problem of the re-normalization of $E(B-V)$ has
been discussed in Mathis (1990; Sect.\,2.1.1).

\section{\bfseries{\itshape SEST} measurements}

\subsection{Observations}

The CO observations toward LH 10:3120 have been carried out
in July 1997 using the 15-m Swedish-ESO Submillimetre Telescope ({\it SEST})
at La Silla, Chile.
A SIS receiver at $\lambda$ = 3\,mm range was employed which yielded
overall system temperatures, including sky noise, of order $T_{\rm sys} =
400$\,K on a main beam brightness temperature ($T_{\rm MB}$) scale.
The backend was an acousto-optical spectrometer (AOS) with 1000 contiguous
channels and the channel separation of 43\,kHz corresponds to
0.11\,km\,s$^{-1}$ at 115 GHz.
The antenna beamwidth was 43\arcsec\ at the observed line
frequencies (Lovas 1992).
The observations were carried out in a dual beam-switching mode (switching
frequency 6\,Hz) with a beam throw of 11\arcmin 40\arcsec\ in azimuth.
All spectral intensities were converted to a $T_{\rm MB}$ scale,
correcting for a main beam efficiency of 0.68 at 115 GHz.
Calibration was checked by monitoring on Orion KL and M17SW and was found
to be consistent between different observation periods within $\pm$\,10\%.
The pointing accuracy, obtained from measurements of the SiO masers
R\,Dor, was better than 10\arcsec.

\subsection{Results}

Fig. 2 shows the results obtained with {\it SEST}
in the direction of LH 10:3120.
The presence of CO near LH 10 already had been
shown by Cohen et al. (1988).
The {\it SEST} contour map (Fig. 3, left panel) shows the presence of $^{12}$CO 
emission over the whole measured area. 
The amount of CO increases to the
north, where the (geometrical) 
centre of the young association LH 10 is located.
The peak of
the CO emission toward LH 10:3120 is near $V_{\rm LSR}=+286$ km\,s$^{-1}$ (Fig.\,3, right panel).
No CO is visible near $+255$ km\,s$^{-1}$, where
the H$_2$ absorption has been found.
Thus, the CO emission takes place at a radial velocity 
$30$ km\,s$^{-1}$ larger than where the 
H$_2$ cloud has its absorption (see Sect.\,2.4).
For the component at $+286$ km\,s$^{-1}$ 
we find a brightness temperature of $T_{\rm MB} = 1.66$ K.
The velocity shift between CO emission and H$_2$ absorption
as well as the lack of CO absorption at LMC velocities
implies that the CO emission near $+286$ km\,s$^{-1}$ 
most likely comes from gas
{\it behind} LH 10:3120.
The {\it SEST} data therefore can not be used to 
compare the CO emission in the direction toward
LH 10:3120 with the H$_2$ column density along this
line of sight. Consequently, the determination of a $X$ factor
for the LMC gas is not possible.

\section{Discussion}

The main results of our measurements of the 
LMC gas towards LH 10:3120 have been compiled
in Table 1. These results allow us to investigate
the correlations between atomic gas, dust and molecules
in the LMC.

Generally, the amount of molecular species is small
in comparison with the total gas quantity.
The hydrogen fraction in molecular form in the
LMC is only $f=2N($H$_2)/[N($H\,{\sc i}$) + 2N($H$_2)$]$
= 0.007$,
lower than typical fractions for
Milky Way gas with similar values in $N$(H\,{\sc i}) 
(Savage et al. 1977). 
Moreover, the upper limit for the CO column density in comparison
to $N$(H\,{\sc i}) is,
with regard to results for the Milky Way 
(Federman et al. 1980), relatively low as well. 
In contrast,
the lower limit of $N$(H$_2$)/$N$(CO)
is $2.0 \times 10^{5}$ and thus
not substantially different from ratios
measured in the Galaxy.
The excitation temperatures for the H$_2$
are 50 K for $J \le 1$ and 470 K for $2 \le J \le 4$ 
(Paper\,I) and do not indicate
an abnormally strong UV radiation 
field in the LMC gas towards LH 10:3120.
A strong UV flux would decrease the molecular gas fraction
through enhanced dissociation.
We thus conclude that the diffuse LMC gas in the line of sight
to LH\,10:3120 initially has a lower molecule content than
comparable gas in the Milky Way.
Recently, low molecule fractions for 
LMC gas also have been proposed
by Gunderson et al. (1998) for two other lines
of sight to the LMC, using spectra at low dispersion
from the {\it HUT} telescope.
Moreover, the only
FUV measurement of H$_2$ in the SMC so far shows a very
low fraction of hydrogen in molecular form, too
(Richter et al. 1998).

\begin{table}[t]
\caption[]{The LMC gas towards LH 10:3120}
\begin{tabular}{rll}
%\hline\noalign{\smallskip}
% & LH 10:3120 \\
%level $J$ & [cm$^{-2}$] & [km\,s$^{-1}$]   & lines used\\
%\noalign{\smallskip}
\hline\noalign{\smallskip}
$N$(H\,{\sc i}) & ..........  & $2.0 \times 10^{21}$ cm$^{-2}$ \\
$N$(H$_2$)  &  .......... & $6.6 \times 10^{18}$ cm$^{-2}$ \\
$N$(CO) & .......... &$ \le 3.3 \times 10^{13}$ cm$^{-2}$ \\
$E(B-V)_{\rm LMC}$ & ..........  & $0.12$ \\
$N$(H\,{\sc i})/$E(B-V)_{\rm LMC}$\,$^a$ & ..........  & $1.7 \times 10^{22}$ cm$^{-2}$\\
$f$ & ..........  & $0.007$\\
$N$(H$_2$)/$N$(CO) & ..........  & $\ge 2.0 \times 10^{5}$\\
\noalign{\smallskip}
\hline
\end{tabular}

\noindent
$^a$ mean galactic value: $4.8 \times 10^{21}$ cm$^{-2}$ (Bohlin et al. 1978)
\end{table}

In searching for explanations for the low molecule fraction
the lower dust content in the LMC gas, as reviewed
by Koornneef (1984), might play a important role.

The gas-to-dust ratio in the LMC gas towards LH 10:3120 is 
with $N$(H\,{\sc i})/$E(B-V) = 1.7 \times 10^{22}$ cm$^{-2}$
about 4 times higher than the galactic value, while
the ratios of $N$(H$_2$)/$E(B-V)$ and
$N$(CO)/$E(B-V)$ are not significantly
different compared to what we find in the Milky
(see Savage et al. 1977; Federman et al. 1980).
Clayton \& Martin (1985) suggested that 
the ratio of $N$(CNO) to $E(B-V)$ might be the same
in the Milky Way and in the LMC, implying that
the amount of dust and the abundance of gaseous CNO elements
are decreased in the same degree.
Does the lower dust content in the LMC (and SMC) also limit 
the amount of molecular hydrogen, at least in the diffuse ISM\,?

Yet, the number of relevant observations is too low to
draw meaningful conclusions about the molecular content 
in the overall diffuse ISM of the Magellanic
Clouds. 
Nevertheless, all measurements made so far 
and discussed here support the scenario, in which
the amount of H$_2$ in the diffuse ISM of the 
Magellanic Clouds is limited by the lower 
dust abundance.
Clearly, the lower metallicity of the Magellanic Clouds 
leads to a smaller dust amount
as well as to a reduced abundance of CO, so that
the lower dust content probably allows (in the mean) less H$_2$ to form.
Through the deficiency of dust 
the shielding from the UV radiation is reduced, too.
Therefore, the total abundance of H$_2$ in the diffuse ISM compared
to the amount of neutral hydrogen is probably lower than
in the Milky Way.
The intrinsic ratio of H$_2$ to CO in the diffuse gas
of a metal-poor galaxy like the Magellanic Clouds
might not be different at all from the one in
the Milky Way, since it is possible that
the abundance of both H$_2$ and CO
is reduced by a similar factor.

For the densest regions in the interstellar gas, the
molecular clouds, the H$_2$/CO ratio is an even 
more critical parameter, since we do not know how different
the cloud sizes in H$_2$ and CO in a low metallicity
environment really are.
Pak et al. (1998) argued that the lower abundances of carbon
and oxygen in the Magellanic Clouds may result in
signifcantly lower cloud sizes as observed in CO emission
compared to the size in molecular hydrogen.
Hence
we might underestimate the total mass of a molecular
cloud when obtaining the H$_2$ amount in a metal-deficient
galaxy using a standard conversion factor between
H$_2$ column density and CO luminosity (Maloney \& Black 1988).
We can not measure the H$_2$/CO ratio {\it within}
the cloud cores. Thus, it is important to find its
value at least in the more diffuse cloud envelopes.  
We then will be able to estimate how certain the
usage of a conversion between CO luminosity and
H$_2$ column density in a galaxy with a low
metallicity can be at all.

Future UV satellites, such as {\it FUSE}, will help to
study the relations between atomic and molecular 
gas in the Magellanic Clouds and other galaxies 
in more detail.
Only good statistics of sight-line measurements
in these systems  will allow a meaningful comparison 
between gas properties 
in metal-deficient galaxies and the Milky Way.

\acknowledgements
We thank John Black for helpful comments and the Heidelberg-T\"ubingen team 
for their great support. 
PR is supported by a grant from the DARA (now DLR) under code 50 QV 9701 3.
YNC thanks the National Science Council of Taiwan for its financial support
through grant \hbox{86-2112-M001-032}.

{}

\begin{thebibliography}{}

\bibitem[]{}
Appenzeller, I., Krautter J., Mandel H., Oestreicher R., 1988, in
`A decade of UV astronomy with {\it IUE}', ESA SP-281, 2, 337

\bibitem[]{}
Barnstedt J., Kappelmann N., Appenzeller I., et al., 1999, A\&AS 134, 561

\bibitem[]{}
Bessel M.S., 1990, A\&A 242, L17

\bibitem[]{}
Bloemen J.B.G.M., Strong A.W., Blitz L., et al., 1986, A\&A 154, 25

\bibitem[]{}
Bohlin R.C., Savage B.D., Drake J.F., 1978, ApJ 224, 132

\bibitem[]{}
Cardelli J.A., Clayton G.C., Mathis J.S., 1989, ApJ 345, 245

\bibitem[]{}
Chin Y.-N., Henkel C., Whiteoak J.B., et al., 1997, A\&A 317, 548

\bibitem[]{}
Chin Y.-N., Henkel C., Millar T.J., et al.,
       1998, A\&A 330, 901

\bibitem[]{}
Clayton G., Martin P.G., 1985, ApJ 288, 558

\bibitem[]{}
Cohen R.S., Dame T.M., Garay G., et al., 1988, ApJ 331, L 95

\bibitem[]{}
de Boer K.S., Jura M.A., Shull J.M., 1987, in 'Exploring the
        Universe at Ultraviolet Wavelengths', eds. Y.Kondo et al.;
        Reidel, Dordrecht, p. 485

\bibitem[]{}
de Boer K.S., Richter P., Bomans D.J., Heithausen A.,  Koornneef J.,
        1998, A\&A 338, L5 (Paper\,I)

\bibitem[]{}
Federman S.R., Glassgold A.E., Jenkins E.B., Shaya E.J.,
       1980, ApJ 242, 545

\bibitem[1996]{}
Fligge M., Solanki S.K., 1997,  A\&AS 124, 579

\bibitem[]{}
Gunderson K.S., Clayton G.C., Green J.C., 1998, PASP 110, 60

\bibitem[]{}
Heithausen A., Mebold U., 1989, A\&A 214, 347

\bibitem[]{}
Israel F.P. 1997, A\&A 328, 471

\bibitem[]{}
Jenkins E.B., Drake J.F., Morton D.C., et al., 1973, ApJ 181, L122

\bibitem[]{}
Koornneef J., 1982, A\&A 107, 247

\bibitem[]{}
Koornneef J., 1984, in IAU Symp.108, 'Structure and Evolution of the
        Magellanic Clouds', eds. S.\,van den Bergh \& K.S.\,de Boer;
        Reidel, Dordrecht, p. 333

\bibitem[]{}
Lovas F.J., 1992, J. Phys. Chem. Ref. Data 21, 181

\bibitem[]{}
Maloney P., Black J., 1988, ApJ 325, 389

\bibitem[]{}
Mathis J.S., 1990, ARA\&A 28, 37

\bibitem[]{}
McGee R.X., Newton L.M., Morton D.C., 1983, MNRAS 205, 1191

\bibitem[]{}
Morton D.C., 1991, ApJS 77, 119

\bibitem[]{}
Morton D.C., Dinerstein H.L., 1976, ApJ 204, 1

\bibitem[]{}
Morton D.C., Noreau L., 1994, ApJS 95, 301

\bibitem[]{}
Pak S., Jaffe D.T., van Dishoeck E.F., Johansson L.E.B., Booth R.S., 1998, ApJ 498, 735

\bibitem[]{}
Parker J.W., Garmany C.D., Massey P., Walborn N.R., 1992, AJ 103, 1205

\bibitem[]{}
Richter P., Widmann H., de Boer K.S., et al., 1998, A\&A 338, L9

\bibitem[]{}
Rubio M., Lequeux J., Boulanger F., et al., 
       1993, A\&A 271, 1

\bibitem[]{}
Savage B.D., de Boer K.S., 1979, ApJ 230, L77

\bibitem[]{}
Savage B.D., Bohlin R.C., Drake J.F., Budich W., 1977, ApJ 216, 291

\end{thebibliography}
\end{document}